\documentclass[reprint,prd]{revtex4-2}
\usepackage[utf8]{inputenc}
\usepackage [english] {babel} 
\usepackage{amsmath,amssymb,latexsym}
\usepackage{slashed}
\usepackage{graphicx}
\usepackage{epstopdf}
\usepackage{mathtools}

\usepackage{tikz}

\usepackage{xcolor}
\usepackage{hyperref}

\def \sec{\begin{section}}
\def \esec{\end{section}}

\renewcommand{\tilde}{\widetilde}
\renewcommand{\bar}{\overline}

\def \th {\theta}

\def \vep {\varepsilon}

\def \Oc {\mathcal{O}}

\def \Lc {\mathcal{L}}

\def \pr {\partial}

\def \oh {\cfrac{1}{2}}

\def \beq { \begin{equation}}
\def \eeq {\end{equation}}

\def\bal#1\eal{\begin{align}#1\end{align}}

\DeclareMathOperator*{\Tr}{Tr}

\def \l {\left}
\def \r {\right}

\def \bra {\langle}
\def \ket {\rangle}

\begin{document}

\title{All holographic systems have scar states}

\author{Alexey~Milekhin}
\affiliation{UC Santa Barbara, Physics Department, Broida Hall, Santa Barbara, CA 93106}
\email{milekhin@ucsb.edu, maybe.alexey@gmail.com}
\author{Nikolay~Sukhov}
\affiliation{Princeton University, Physics Department, Jadwin Hall, Princeton, NJ 08540}
\date{\today}

\begin{abstract}
    Scar states are special finite-energy density, but non-thermal states of chaotic Hamiltonians. We argue that all holographic quantum field theories, including $\mathcal{N}=4$ super Yang--Mills, have scar states. Their presence is tied to the existence of non-topological, horizonless soliton solutions in gravity: oscillons and a novel family of excited boson stars. We demonstrate that these solutions have periodic oscillations in the correlation functions and posses low-entanglement entropy as expected for scar states. Also we find that they can be very easily prepared with Euclidean path integral.
\end{abstract}

\maketitle

\tableofcontents

\section{Introduction}

\begin{figure}
    \centering
    \includegraphics{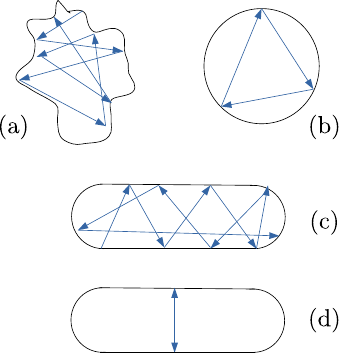}
    \caption{Illustration of possible one-particle motions in a cavity. (a) is a fully chaotic cavity, (b) is integrable circular cavity. The cavity in (c) and (d) is the Bunimovich stadium. Case (d) is the scar trajectory. }
    \label{fig:scar}
\end{figure}

Everyday experience tells us that most systems thermalize over time, but there are exceptions. For instance, consider a single classical particle moving inside a reflecting cavity - Figure \ref{fig:scar}. For a cavity of random shape,  all trajectories will look essentially random and they will explore of all the phase space - Figure \ref{fig:scar}~(a). This is a one-particle counterpart of thermalization: time average along such trajectory will be equal to the phase space average. We may call such cavity ergodic. Of course, there are special (integrable) shapes for which all trajectories have a short period of oscillations - Figure \ref{fig:scar}~(b). %However, such situations are %rare and, in general, not %stable to perturbations.
However, there are intriguing cases like the Bunimovich stadium\cite{bunimovich1979ergodic}: a generic trajectory is thermal - Figure \ref{fig:scar}~(c), but there is a set of short periodic trajectories which do not explore all of the phase space - Figure \ref{fig:scar}~(d). We will refer to them as \textit{classical scars}.

There is an exponentially small number of such trajectories and they are unstable, so one might expect that on a quantum level they are not important for anything. \textit{It turns out to be incorrect} \cite{PhysRevLett.53.1515}. A number of energy eigenstate wavefunctions are concentrated around the classical scar trajectories. In other words, short classical unstable orbits permanently "scar" the wavefunctions. This is the phenomena of \textit{single-particle quantum scars}.

 Recent interest towards scars started from discovering a similar phenomena experimentally in a \textit{many-body quantum system}  of cold Rydberg atoms \cite{Bernien_2017,turner2018weak}(see also \cite{Su_2023}).
In the many-body setup, there is no classical analogue and the scarring occurs in the Hilbert space: evolution of certain states $| \Psi(0) \ket$ shows short-period revivals when the fidelity $|\bra \Psi(t) | \Psi(0) \ket| \approx 1$.
 It is important to emphasize that  scars were observed in ergodic Hamiltonians, for which most of the states are thermalizing: the fidelity exponentially decays to zero and it stays zero till the Poincare recurrence time.
 A hallmark of many-body quantum scars are scarred energy eigenstates with abnormally low entanglement compared to states of the same energy density. 
Typically they represent an exponentially small fraction of the Hilbert space.
 We refer to \cite{serbyn2021quantum,Moudgalya:2021xlu,Papic2022,Chandran:2022jtd} for an overview.

The main objective of this paper is to explore the scarring phenomena in gravity and quantum field theory.
AdS/CFT correspondence \cite{Susskind:1994vu,Maldacena:1997re,Gubser:1998bc,Witten:1998qj} says that certain strongly-coupled large $N$ conformal field theories (CFTs) are dual to classical gravitational theories in the asymptotically anti de-Sitter (AdS) spacetime. The gravitational dual of thermalization process is the formation of a black hole. It is known \cite{Bizon:2011gg,Garfinkle:2011hm,Jalmuzna:2011qw,Buchel:2012uh,Dias:2011ss} that $AdS$ spacetime is unstable: a small perturbation (either matter or metric) inside $AdS$ quickly collapses into a black hole. However, there are special initial conditions which do not lead to a collapse. Instead the perturbation oscillates inside $AdS$ forever \cite{Maliborski:2013jca,Horowitz:2014hja}. This is a clear analogue of a classical scar trajectory.
It is important to emphasize that these oscillating gravity solutions and "small" perturbations are large from the boundary CFT point of view, they correspond to the energy density of the order of the central charge (in the units of boundary volume). The simplest example, which we study in the present paper, involves Einstein gravity
minimally coupled to a matter scalar field.

Given that there are classical scar solutions in classical gravity, one \footnote{We are indebted to F.~Popov for asking the first question and explaining to us the difference between the various scar phenomena.} can naturally ask two question:
\begin{itemize}
    \item Is there associated quantum scarring phenomena in the wavefunctions of quantum gravity? 
    \item What is the CFT state dual to classical gravity scar?
\end{itemize}
In this paper we would like to answer the second question:

\textit{Such periodic classical gravity solutions, known as boson stars or oscillons, are holographically dual to many-body scar states at the boundary. Moreover, we would like to argue that it is a feature of all holographic systems, whenever the bulk has 3 or mode spacetime dimensions and has a scalar field. One such example is $\mathcal{N}=4$ super Yang--Mills which we will discuss in detail.}

Why is this interesting? Holographic systems are supposed to be not just chaotic \cite{Hayden:2007cs,Sekino:2008he}, but maximally chaotic \cite{Shenker:2013pqa,Shenker:2013yza,Maldacena:2015waa}. Having scar states is not a generic feature of chaotic Hamiltonians. However, our result indicates that scars, which break erodicity, are generic for holographic systems. Also, it is expected that the presence of scars is associated with hidden symmetries (or more generally, spectrum-generating algebras) \cite{Bu_a_2019,Medenjak_2020,Lin_2019,Mark_2020,Ren:2020dcs,Pakrouski:2021jon,Ren:2021khc,Moudgalya:2022nll,Sun:2022oew}.
The question of a possible hidden symmetry
behind oscillons has been extensively discussed in the literature before and we give a small overview in the Conclusion.
Also there is an interesting difference with classical scars.
It is known that boson stars and oscillons are
linearly stable and exhibit slow thermalization: adding an extra perturbation on top does not immediately lead to black hole formation \cite{Dias:2012tq}. In contrast, classical scars are associated with unstable periodic orbits.
%It would be very interesting to understand the origin of this symmetry separately within gravity and CFT.
Finally, the presence of boson stars and oscillons gives predictions for certain boundary CFT correlation functions involving a non-primary operator $e^{\vep \Oc}$, $\Oc$ being the single-trace CFT scalar, dual to the 
scalar field in the bulk.

Oscillons and boson stars require scalar matter fields. Even more generally, there are geon solutions \cite{Horowitz:2014hja} which are made purely from the metric, that is, from the CFT stress-energy tensor. However, they are more complicated and they break translational symmetry at the boundary so we do not consider them here.

\begin{figure}
    \centering
    \includegraphics[scale=0.5]{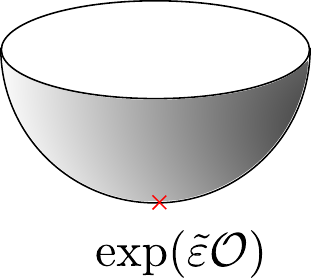}
    \caption{Euclidean path-integral state preparation of oscillons and boson stars: we put CFT on a Euclidean hemisphere and insert $e^{\tilde{\vep} \Oc}$ at the south pole.}
    \label{fig:eucl_prep}
\end{figure}

How hard is it to prepare these states?
The main difference between oscillons and boson stars is whether the scalar field is real or complex: for oscillons the field is real and the metric is time-dependent. For boson stars the field is complex and has a harmonic time-dependence $e^{-i \Omega t}$, so the stress-energy tensor and the metric are time-independent.
From the gravity point of view, oscillons and boson stars are very different solutions which are usually discussed separately. Interestingly, we find that from the boundary CFT viewpoint they are very similar and both of them can be very easily prepared using the Euclidean path integral on a hemisphere with a single operator insertion $e^{\tilde{\vep} \Oc}$ at the pole - Figure \ref{fig:eucl_prep}.  In case of oscillons the single-trace operator $\Oc$ is hermitian, whereas for boson stars it is complex. In our regime of interest, $\tilde{\varepsilon}$ can be large, of order the square-root of central charge if the two-point function of $\Oc$ is normalized to $1$. The resulting CFT state lives on sphere $S^{d-1}$.  We find evidence that for boson stars  it is possible to take the infinite volume limit to get a homogeneous state on $\mathbb{R}^{d-1}$. For oscillons we were not able to find such limit.

It is important to emphasize that oscillons and boson stars are not energy eigenstates for the boundary CFT. 
However, they are very close to being energy eigenstates: their energy density scales with the (large) central charge, whereas the variance is expected to scale at most as the square-root of the central charge,  essentially because the classical gravitational solution provides a dominant saddle-point for the path-integral. 
Nonetheless, they are non-thermalizing, uniform energy density, pure
states which support 
eternal oscillations in the 1-point function of a scalar single-trace operator $\Oc$:
\begin{align}
\label{eq:expvalue}
& \text{oscillon:} \ \bra \Psi | \Oc | \Psi \ket = \sum_{i=1} \vep_i \cos((2i-1)\Omega t), \\ \nonumber
& \text{boson star:} \ \bra \Psi | \Oc | \Psi \ket = \vep e^{i \Omega t}.
\end{align}
Such behavior, by definition, implies the violation of the eigenstate thermalization hypothesis (ETH) \cite{Srednicki:1994mfb,PhysRevA.43.2046}, which only allows such oscillating
terms to be exponentially small in the thermal entropy.
More interestingly, it was recently argued \cite{Alhambra:2019xaa} that under certain mild assumptions in discrete local systems, \textit{the presence of such revivals in pure states possessing area-law entanglement imply the presence of scarred energy eigenstates in the spectrum.} In this paper we indeed find evidence that boson stars possess area law entanglement at the boundary: entanglement of a CFT subregion scales as the area of the boundary of that region.
 In contrast, for a thermal (black hole) state the entropy scales as the volume of a CFT subregion.
Hence we can expect the presence of exact scars in the spectrum. The simplest oscillons we study here only exist in a finite volume and up to some critical energy density, so it is not clear how to separate volume and area law entanglement for them.

Let us summarize out key findings:
\begin{itemize}
    \item Boson stars and oscillons can be easily prepared with Euclidean path integral.
    \item They have lower boundary entanglement entropy compared to a black hole of the same mass. In case of boson star this separation is parametric: area law instead of volume law. 
    \item We find that oscillons have bounded mass. However, we uncover a novel family of linearly-stable excited boson stars, which we call \textit{C-stars}, for which the mass is unbounded. 
\end{itemize}
In Section \ref{sec:BS} we will explain why one needs to study excited boson stars, rather than
fundamental (non-excited ones) to find scars.

Recently the phenomena of scar states in quantum field theories (QFT) has been addressed in a number of papers.
Previous discussions of oscillons and boson stars within
the AdS/CFT include \cite{Buchel:2012uh, Buchel:2013uba, Buchel:2014dba,
Balasubramanian:2014cja,Buchel:2015rwa,
Craps:2021xmk}.
Scars based on Virasoro symmetry and their relation to $AdS_3/CFT_2$ are discussed in \cite{Caputa:2022zsr,Liska:2022vrd}.
For a general discussion of scar states within the QFT framework we refer to \cite{Cotler:2022syr, Desaules:2022ibp,Delacretaz:2022ojg}.
Another recent discussion of scar states \cite{Dodelson:2022eiz} is based on stable orbits around black holes. 

The rest of the paper is organized as follows. Section \ref{sec:oscillons} is dedicated to oscillon solutions.
We discuss their generic properties and then switch to the entanglement entropy in Section \ref{sec:ee}. In Section \ref{sec:n4} we argue that oscillons exist in the supergravity dual to $\mathcal{N}=4$ super Yang--Mills.
Section \ref{sec:BS} is devoted to boson stars. We briefly describe their properties and demonstrate
that they have area-law entanglement. 
The CFT state preparation of oscillons and boson stars is discussed in Section \ref{sec:prep}.
In Section \ref{sec:weak_energy} we turn away from discussing
specific solutions and argue generally that black holes maximize entanglement entropy due to weak energy condition.
In Conclusion we summarize our findings and outline open question.

\section{Oscillons}
\label{sec:oscillons}
In this Section we study the oscillons states first found in \cite{Maliborski:2013jca}. We will review their perturbative construction and then compute subsystem entanglement entropy.

Consider Einstein gravity with negative cosmological constant plus a minimally-coupled scalar field $\phi$, which is spherically symmetric. The main statement of \cite{Maliborski:2013jca} is that such soliton can exist forever, without collapsing into a black hole. The Lagrangian is
\beq
\Lc = \frac{1}{16 \pi G_N} (R + 2 \Lambda) - (\pr_\mu \phi)^2 - m^2 \phi^2.
\eeq
A general ansatz for the metric, preserving the spherical symmetry is
\beq
\label{eq:geon_metric}
ds^2 = \frac{l^2}{\cos^2 x} 
\l( -dt^2 A e^{-2 \delta} + A^{-1} dx^2 + \sin^2 x d\Omega_{d-1}^2 \r).
\eeq
Usual (undeformed) $AdS_{d+1}$ is $A(x,t)=1, \delta(x,t)=0$. $AdS$ radius $l$ is determined by $l^2 = \frac{d(d+1)}{2\Lambda}$.
The boundary is at $x=\pi/2$ and the center is at $x=0$. We impose a gauge constraint that at the boundary $\delta$ is zero: $\delta(t,\pi/2)=0$, so the dimensionless coordinate $t$ is the boundary time. This setup corresponds to $(d-1)$+1-dimensional CFT located at the asymptotic boundary $S^{d-1} \times \mathbb{R}_t$.

It is very important to discuss units in this paper because all results we present will be in dimensionless units. The conformal metric at the boundary has unit radius. Correspondingly, $t$ and the frequencies are measured in the units of the boundary radius. 
$AdS$ radius $l$ drops out from the equations and we can reabsorb $8 \pi G_N = l_p^{d-1}$ into $\phi$. So the scalar field is measured in the units of $l_p^{-(d-1)/2}$. With Dirichlet boundary conditions for the scalar field, function $A$ has the following expansion:
\beq
A = 1 - 2 M (\pi/2-x)^d + \dots
\eeq
CFT energy density $T_{tt}$ is proportional to $M$ times $(l/l_p)^{d-1}$. The ratio $(l/l_p)^{d-1}$ is proportional to the CFT central charge, which is large. Similarly, using RT/HRT prescription \cite{Ryu:2006bv,Hubeny:2007xt}, entanglement entropy of boundary subregions is given by the area of extremal co-dimension two surfaces in the bulk with minimal area:
\beq
S_{vN} = \frac{\text{Area}}{4 G_N}.
\eeq
Technically it is always infinite, because $AdS$ boundary is infinitely far, so we will always compute the difference with the vacuum (empty $AdS$ answer).
So in this paper we compute it in the units of $(l/l_p)^{d-1}$. The upshot is that we are interested in large CFT perturbations, when the energy density and entropy are proportional to the central charge.
The solutions are parametrized only by the (dimensionless) value of the scalar field (in the $l_p^{-(d-1)/2}$ units).

Without gravitational backreaction, a minimally coupled scalar of mass $m^2 = \Delta(\Delta-d)$ has a set of (spherically symmetric) normal modes:
\begin{align}
 \label{eq:normalModes}
e_j(x) = n_j \cos(x)^\Delta P_j^{d/2-1,\Delta-d/2}(\cos(2x)), \\ \nonumber
 j=0,1,\dots, \\ \nonumber
 n_j^2 = \frac{2(2j+\Delta)\Gamma(j+\Delta) j!}{\Gamma(j+d/2)\Gamma(j+\Delta-d/2+1)},
\end{align}
with frequencies $\omega_j = \Delta + 2j$ and
$P$ being the Jacobi polynomials. With this choice of normalization, they are orthonormal with respect to the $\tan(x)^{d-1}$. The fundamental mode $j=0$ has no zeroes, and the excited ones have $j$ zeroes.
The question is what happens with this solution once we include backreaction into account.
\textit{The key statement is that solutions with a single dominant frequency do not collapse into a black hole.} 
Solutions which have several frequencies collapse very quickly, that is, they thermalize. 
Let us list a few other facts:
\begin{itemize}

    \item Non-spherically symmetric configurations are more prone to instability: only very special modes can be dressed to yield a periodic solution \cite{Dias:2017tjg}.
    \item For generic perturbations (not oscillons),  secularly growing corrections arise in the 3rd order of perturbation theory, $\vep^3 t$. So thermalization time is of order $1/\vep^2$. 

    \item Even if the scalar field has
    zero self-interaction, tree-level graviton exchange induces it. Hence one might expect that explicit self-interaction should not change the picture much. This logic was verified numerically in \cite{Kim:2014ida, Kim:2015rvu, Cai:2015jbs} by considering $\lambda \phi^4$ interaction in the bulk.

    \item Finally, one can ask about the influence of higher-derivative terms in the gravity action. It was argued based on numerical analysis that in Einstein--Gauss--Bonnet
    \footnote{Einstein--Gauss--Bonnet gravity has a mass gap for black holes, making $AdS$ stable again. The claim is that there are initial conditions, above this gap, which do not lead to black hole formation}
    gravity \cite{Buchel:2014dba,Deppe:2014oua,Deppe:2016dcr} and Starobinsky $R^2$ gravity \cite{Zhang:2016kzg}, oscillons continue to exist. 
\end{itemize}
All this suggests that oscillons have lifetime non-perturbative in $1/G_N$. Meaning that their lifetime is $e^{\#/G_N}$, non-perturbatively large in $1/G_N$.

\begin{figure}
    \centering
    \includegraphics[scale=0.9]{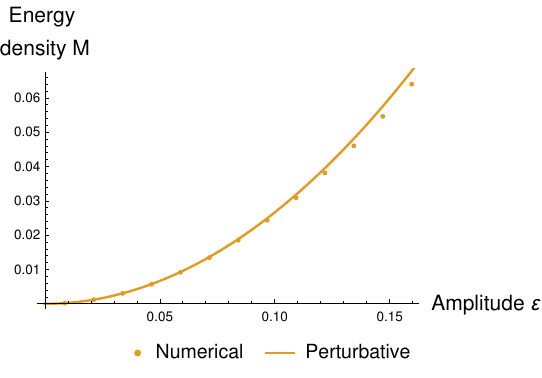}
    \includegraphics[scale=0.9]{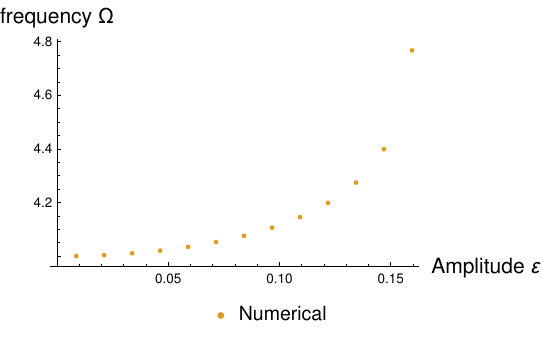}
    \caption{Boundary energy density (in units of $(l/l_p)^{d-1}$) and frequency (in units of boundary radius) of oscillon
    as the function of the amplitude (in units of $l^{-(d-1)/2}_p$).  At finite value of the amplitude the frequency blows up, but the mass stays constant. These plots are made for a massless scalar in asymptotically $AdS_5$ but similar behavior occurs for massive fields and in other dimensions.}
    \label{fig:ads5osc}
\end{figure}

For example, one can start from the lowest mode solution in asymptotically $AdS_5$ spacetime with $j=0, \Omega=4$:
\beq
\phi = \vep e_0(x) \cos(4t),
\eeq
and then try to find the fully backreacted solution:
\beq
\label{eq:osc_ansatz}
\phi = \sum_{i,j} f_{i,j} e_j(x) \cos((2i+1) \Omega t).
\eeq
Functions $e_j(x)$ form a basis, so the only special property of this ansatz is the periodic time-dependence.
By AdS/CFT dictionary, such field profile leads to the oscillating expectation value of the dual operator $\Oc$ in the form (\ref{eq:expvalue}). 
One can add backreaction either perturbatively or perform numerics.
Following the approach of \cite{Maliborski:2013jca}, we constructed such solutions numerically.
In short, one truncates the expansion (\ref{eq:osc_ansatz}) at some big values of $i,j$ and then requires the Einstein equations to be satisfied on a set of collocation points in space and time. We fix the amplitude $\vep$ by requiring $f_{0,0} = \vep$.
The resulting mass $M$ and frequency $\Omega$ for asymptotically $AdS_5$ space and massless scalar field (which is the case relevant for $\mathcal{N}=4$ super Yang--Mills) are shown on Figure \ref{fig:ads5osc}. One distinct
feature we find for various spacetime dimensions and various masses is that the frequency $\Omega$ blows up at a finite value of the scalar field amplitude, whereas the mass stays finite. 

One interesting question is whether for $AdS_3$ these solutions can have mass above the BTZ black hole threshold. For a massless field with Dirichlet boundary conditions the maximal mass appears to be much below.

\subsection{Entanglement entropy}
\label{sec:ee}
Let us discuss the entanglement entropy.
Since the metric is time-dependent the entanglement entropy is expected to be time-dependent too. However, we do not expect it to change a lot during the period of one oscillation, therefore we will concentrate on the time-symmetric $t=0$ slice for which we can use a simple RT prescription.
For small subsystems of linear size $s$ the entanglement entropy grows very slowly:
\beq
S_{vN} \sim \vep^2 s^{d}, \ \vep^2 s^{d} \ll 1. 
\eeq
(as usual, in the units of $(l/l_p)^{d-1}$).
The origin of this equation is the following. Oscillon has a finite energy density $M \sim \bra T_{tt}\ket$ at the boundary, proportional to $\vep^2$. 
The behavior $\bra T_{tt}\ket s^{d}$ for small subsystems was previously proved in \cite{Bhattacharya:2012mi,Faulkner:2013ica}: for small subsystems the RT surface lies near the boundary and the only important parameter in the metric is $M \sim \bra T_{tt}\ket$, $s^d$ comes from dimension analysis.
Of course, this does not imply volume law entanglement or area law, for that we need to look at large subsystems. 

 The problem is that we are dealing with global $AdS$, so the boundary CFT lives on a sphere. But what is a large subsystem of a sphere? Similar problem arises in condensed matter setups because they study systems of finite number of spins.
 In principle, we can do a Weyl transformation to map the state of the CFT from a sphere to a plane. 
 On the gravity side it corresponds to appropriately selecting a Poincare path inside global $AdS$.
 However, the resulting state will be time-dependent and inhomogeneous \cite{Horowitz:1999gf,Nozaki:2013wia}, so it is not very useful. 
 One natural thing to do is to compute the entropy for half of the system. Then we have only two parameters because we have a CFT:  radius of the boundary sphere $r$ (which we set to unity) and energy density $M \sim \bra T_{tt} \ket$. The entanglement entropy depends only on the effective dimensionless length
$r \bra T_{tt} \ket^{1/d} \sim r M^{1/d}$. Then the volume-law entanglement in $d-1$ spacial dimension can be associated with
\beq
\text{volume law:}\ S_{vN} \sim (r M^{1/d})^{d-1} \sim M^{(d-1)/d}
\eeq
growth for large $M$, whereas the area-law is
\beq
\text{area law:}\ S_{vN} \sim (r M^{1/d})^{d-2} \sim M^{(d-2)/d}.
\eeq
This is the same as conventional area and volume law entanglement: we can fix a smaller subsystem of size $s$ and send $M$ to infinity, $s$ to zero, keeping $s^d M$ fixed, but big. In this limit CFT is effectively decompactified and large $s^d M$ governs the entanglement behavior of large subsystems.

Black holes yield volume-law $M^{(d-1)/2}$. This can be understood without any computations: in this regime the horizon lies very close to $x=\pi/2$, namely $x_h \sim \pi/2-1/M^{1/d}$. The RT surface will wrap around the horizon and most of its length will come from a disk $x=x_h$, which area is proportional to $1/\cos^{d-1}(x_h) \sim M^{(d-1)/d}$.

\begin{figure}
    \centering
\includegraphics[scale=0.9]{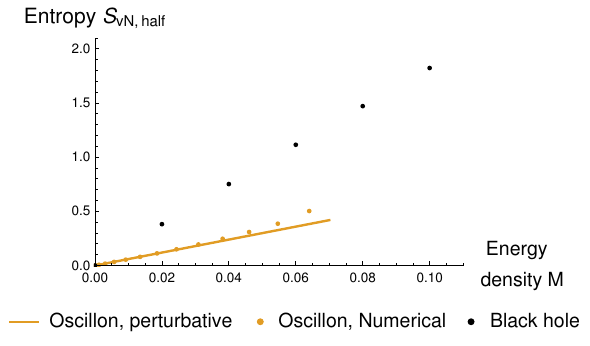}
\caption{Entanglement entropy (again, in units of $(l/l_p)^{d-1}$) of half the system for $AdS_5$ black hole and oscillon.}
    \label{fig:oscillon_EE}
\end{figure}

Unfortunately, for oscillons $M$ has a maximum value, so we cannot distinguish the area law and the volume law. The only thing we can verify is that oscillons have lower entanglement entropy, compared to black holes. This is indeed the case as illustrated by Figure \ref{fig:oscillon_EE}. 
In the next Section we will study boson stars for which the mass $M$ can be unbounded. We will see that they indeed exhibit area law.

\subsection{A comment on $\mathcal{N}=4$ super Yang--Mills}
\label{sec:n4}
$\mathcal{N}=4$ super Yang--Mills is dual to $AdS_5 \times S^5$ solution in IIB ten-dimensional supergravity. This background is sourced by (self-dual) 4-form field $H$. We would like to claim that there exist oscillons in this background which only propagate along $AdS_5$ part. Meaning that this solution keeps the radius of $S^5$ constant. The relevant oscillating scalar field is the dilaton $\phi$ or the axion $\chi$.
In the Einstein frame the Lagrangian looks like
\beq
\frac{1}{16 \pi G^{(10)}_{N}} \l( R - \oh(\pr \phi)^2 - \oh e^{ 2\phi} (\pr \chi)^2 - \frac{1}{4} (dH)^2 \r),
\eeq
where $G^{(10)}_N$ is ten-dimensional Newton constant. 

There is non-trivial flux of $H$ through $S^5$, but the dilaton and axion are constant. Since $H$ has traceless stress-energy tensor, Einstein equations with non-constant $\phi$ and $\chi$ can be written as
\beq
R_{\mu \nu} =  \pr_\mu \phi \pr_\nu \phi + e^{2 \phi} \pr_\mu \chi \pr_\nu \chi + T^H_{\mu \nu} .
\eeq
Hence, if $\phi$ and $\chi$ do not vary over the $S^5$, one can make an ansatz for $AdS_5$ deformation like eq. (\ref{eq:geon_metric}), but with $S^5$ having a constant radius.

In these units (which are different from the rest of the paper) dilaton $\phi$ is dimensionless, the corresponding dual operator is $\Tr F^2/g_{YM}^2$, 
its two-point function is proportional to $N^2$ ($F$ is the gauge field strength). 
In Section \ref{sec:prep} we will discuss the state preparation.
In order to produce order 1 correction to the metric, the operator insertion at the Euclidean disk should have the form $\exp\l( \tilde{\vep} \frac{1}{g_{YM}^2}  \Tr F^2\r)$, where $\tilde{\vep}$ is order 1 small number (say, 0.01).

Strictly speaking\footnote{We are grateful to A.~Buchel for a discussion on this point.}, the stability of oscillons has been shown only for the $AdS$ metric and the scalar field perturbations. In the linear regime the $S^5$ part will add extra scalar fields charged under $SO(6)$ symmetry. 
We do not expect that these extra fields will destabilize the oscillon. For empty $AdS$ the extra fields have normal modes with frequencies away from zero and for small oscillon amplitudes the shift of the normal mode frequencies will be small. But it would be instructive to study this question more carefully.

\section{Boson stars}
\label{sec:BS}

In this Section we make a minimal modification and study a "phenomenological" holographic model: Einstein--Maxwell theory minimally coupled to a complex scalar field. We refer to \cite{Cvetic:1999un,Cvetic:1999xp,Cvetic:2000dm,Cvetic:2000nc,Buchel:2015rwa} for more "realistic" Einstein--Maxwell--scalar theories arising from higher--dimensional supergravities. 
For simplicity, we consider massless scalar in $1+3$ dimensions.
This theory has a plethora of different phases and solutions, including hairy black holes and boson stars \cite{Hu:2012dx,Dias:2011tj,Arias:2016aig}, depending on the value of the charge. In this paper we would like to point out the existence of an extra family of heavy boson stars, which we call \textit{C-stars}. It was first discovered in \cite{Hu:2012dx} but then overlooked in the subsequent works. Here we study their properties in more detail, demonstrate their linear stability and argue that they represent scar states in the dual CFT.

It is expected that there are no global symmetries is quantum gravity, this is why we do not study a complex field with global $U(1)$.
By AdS/CFT correspondence gauged $U(1)$
symmetry in the bulk corresponds to a global symmetry in the CFT. Gauge coupling constant 
is related to the coefficient in the operator product expansion (OPE) of two current operators in the CFT.

The Lagrangian is
\beq
\label{eq:model_lagrangian}
\Lc = \frac{1}{16\pi G_N} (R + 2 \Lambda) - |D_\mu \phi|^2 - \frac{1}{4} F_{\mu \nu} F^{\mu \nu},
\eeq
\beq
D_\mu \phi = \pr_\mu \phi - i e A_\mu \phi. \nonumber
\eeq

\begin{figure*}
    \centering
    \includegraphics[scale=0.8]{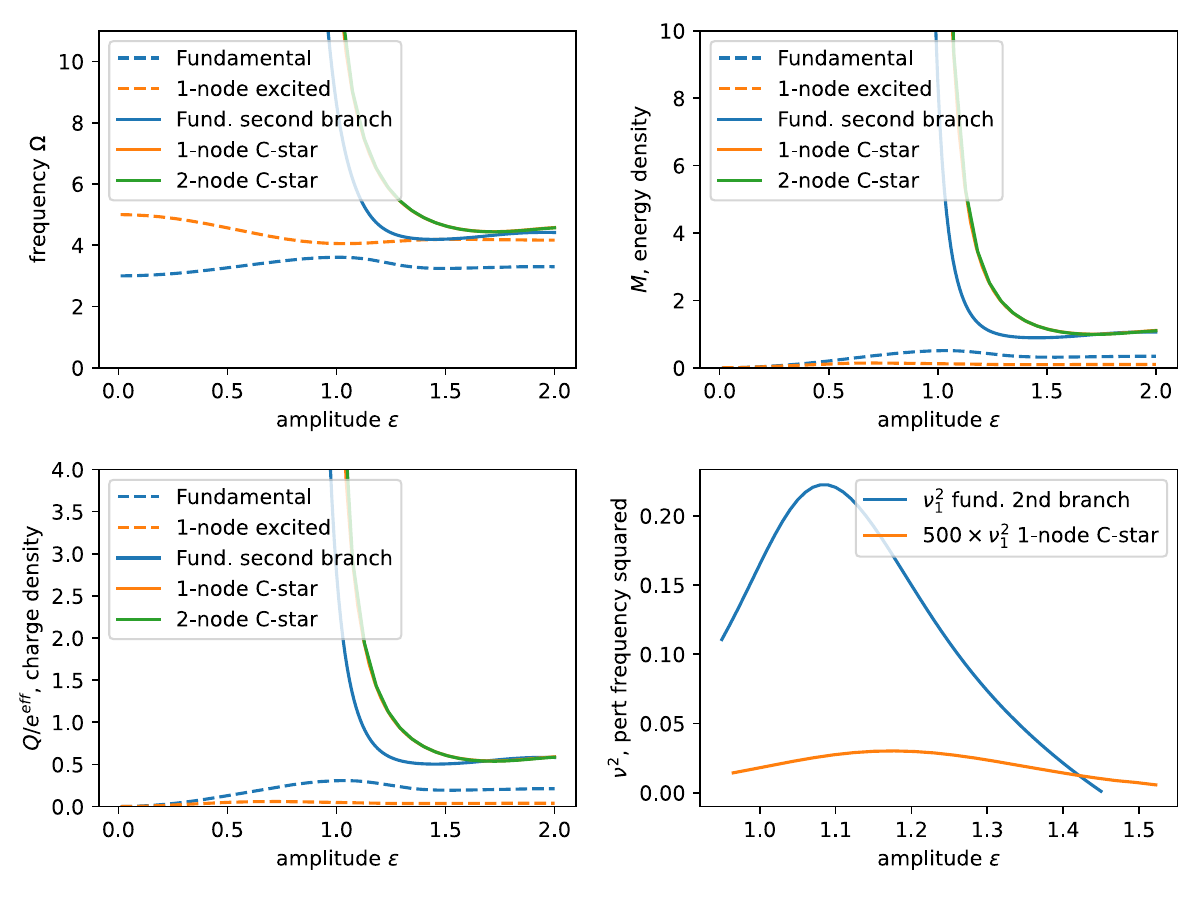}
    \caption{Asymptotically $AdS_4$, intermediate coupling $e^{\rm eff}_{\rm crit,1}<e^{\rm eff}=2.2<e^{\rm eff}_{\rm crit,2}$: frequency $\Omega$ and mass $M$ of fundamental boson stars (blue) and C-stars (orange and green, almost coincide) as a function of the scalar field value at the center $|\phi(0)|=\vep$. Both the fundamental (dashed blue), the second branch fundamental (solid blue) and C-stars (orange, green) has a local maximum in the mass, after which the solutions with bigger amplitudes are linearly unstable.  }
    \label{fig:preC}
\end{figure*}

\begin{figure*}
    \centering
    \includegraphics[scale=0.8]{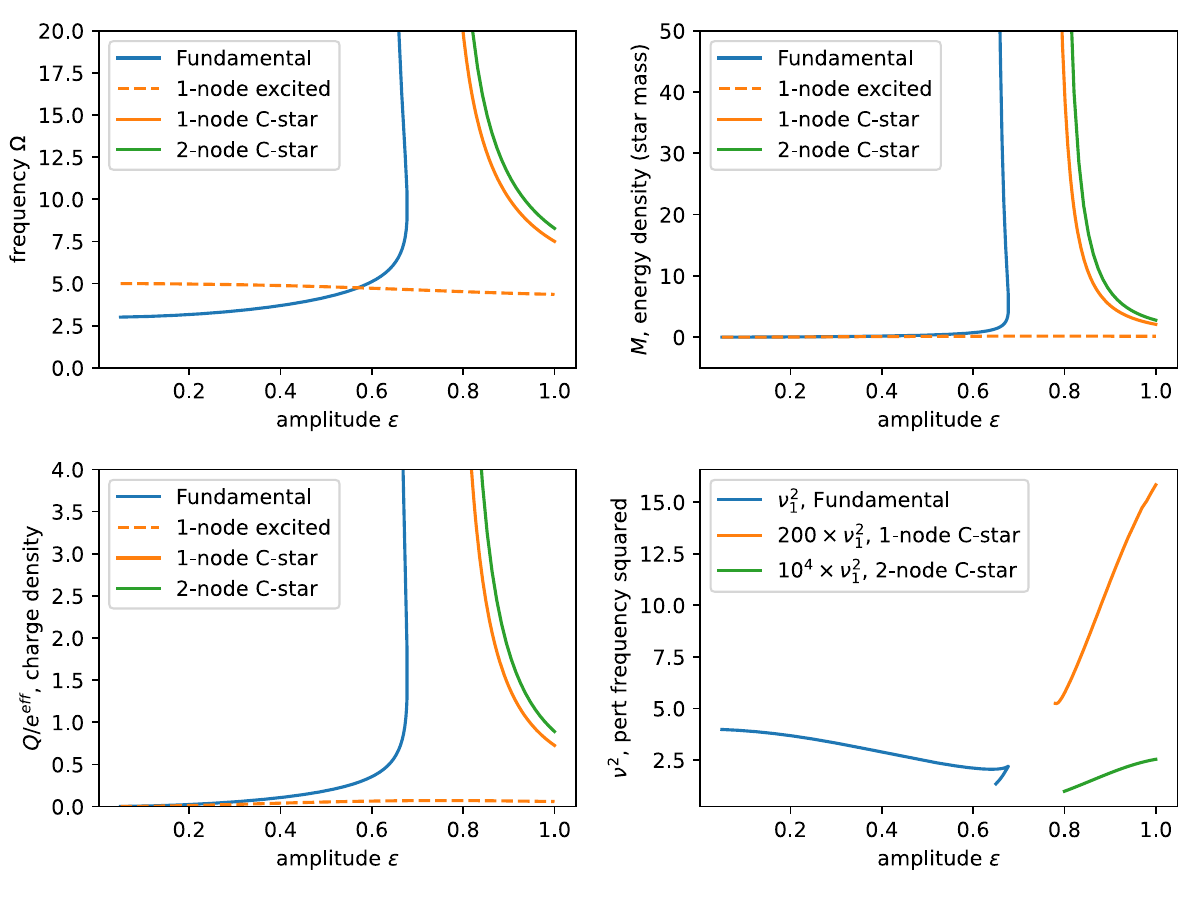}
    \caption{Asymptotically $AdS_4$, strong coupling $e^{\rm eff}=3>e^{\rm eff}_{\rm crit,2}$. Upper panel: frequency $\Omega$, mass $M$. Lower panel: charge $Q$ and the lowest frequency of linearized perturbations.
    Orange and green solid curve is the C-star:
    it is a solution where the scalar field has one (orange) or two (green) zeros, but it is not continuously connected to the normal modes $e_{1,2}(x)$ of the scalar, which is shown in dashed. C-star is linearly stable: its perturbations frequencies are close to zero, but they are real.
    }
    \label{fig:postC}
\end{figure*}

We again put $l_p^{d-1}= 8 \pi G_N=1$ and measure the scalar field in unites of $l_p^{-(d-1)/2}$ and gauge field $A_\mu$ in units of $l l_p^{-(d-1)/2}$.
We can do rescaling of the boson star equations, which reveals that the only important parameter (in addition to the value of the fields) is $e^{\rm eff}=e l/l_p^{(d-1)/2}$. In holography we expect it to be of order $1$, because the interaction strength is of order the gravitational one (suppressed by the CFT central charge).

We again study spherically-symmetric solutions in the form
(\ref{eq:geon_metric}), but the scalar field has a simple one-harmonic time-behavior:
\beq
\phi(t,x) = e^{-i \Omega t} \phi(x).
\eeq
In the limit of vanishing backreaction (very small amplitude), $\phi(x)$ are the normal modes (\ref{eq:normalModes}) inside empty $AdS_{d+1}$. 
Because the stress-energy tensor is proportional to
$|\phi|^2$, the actual metric is time-independent. This is why to find the solutions we can use a simple shooting method. Since the equations of motion for the scalar field are singular both at the origin $x = 0$ and at the AdS boundary $x = \pi/2$, we step away from the origin using power series expansion at the origin, numerically integrate the equations up to a point $x_1$ close to the boundary and then use the scalar field, the gauge field and the metric functions values at $x_1$ to fit an asymptotic power series expansion near the boundary. We then shoot for the scalar field frequency $\Omega$ to match the scalar field derivative $\phi'_l(x_1)$ found by numeric integration to the scalar field derivative found from the asymptotic $\phi'_r(x_1)$. We again impose Dirichlet boundary conditions for the scalar, such that near the boundary $\tilde{\phi} \sim (\pi/2-x)^d$. The only non-zero component of the vector potential is $A_t$.
This component and the metric has the following expansion near the boundary:
\beq
A_t(x) \approx \frac{Q l}{d-2} (\pi/2-x)^{d-2} + \dots,
\eeq
\beq
A(x) \approx 1-2M (\pi/2-x)^d + \dots
\eeq

Before diving inside the details, let us discuss the expectations from the CFT side.
From the CFT perspective such state has non-zero global $U(1)$ charge density $\propto Q$ and 1-point expectation value of a charged operator $\Oc$:
\beq
\bra \Psi | \Oc | \Psi \ket \sim e^{-i \Omega t}.
\eeq
As in the case of  oscillons, we are interested in the "large" masses and electric charges, of the order of the CFT central charge. In this case in a given charge sector, the lowest energy state has non-zero energy density, proportional to the mass $M$. In the limit of large charge $Q$ there are many field-theory results 
relating $Q$ to $M$ \cite{Hellerman:2015nra,Hellerman:2017sur,Jafferis:2017zna,Badel:2019khk,Hellerman:2020sqj,Hellerman:2021yqz,Alvarez-Gaume:2019biu,Cuomo:2020rgt}. Specifically for $2+1$ CFT it is expected that $M \sim Q^{3/2}$ and in $3+1$ dimensions $M \sim Q^{4/3}$.

It would be convenient to explore the space of gravity solutions by fixing the amplitude $|\phi(0)| \equiv \vep$ of the scalar field at the center $x=0$ and the asking what discrete set of frequencies $\Omega$ are allowed. 
In short, there are three phases, depending on the effective charge \cite{Gentle:2011kv,Dias:2011tj,Arias:2016aig} $e_{\rm eff}$.

\begin{figure}
    \centering
\includegraphics[scale=0.9]{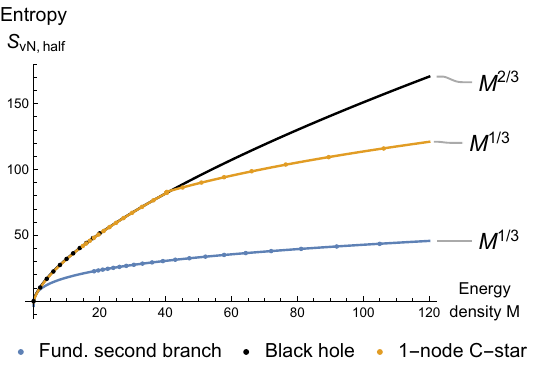}
    \caption{Asymptotically $AdS_4$: boundary EE of half the system for extremal RN black hole, C-star and the (second branch) fundamental boson star in the intermediate regime when $e^{\rm eff}_{\rm crit,1}<e^{\rm eff}=2.2<e^{\rm eff}_{\rm crit,2}$. Dots are numerical data, solid lines represent a fit. The corresponding  phase diagram is shown in Figure \ref{fig:preC}.}
    \label{fig:ads4_precross}
\end{figure}

\begin{figure}
    \centering
\includegraphics[scale=0.9]{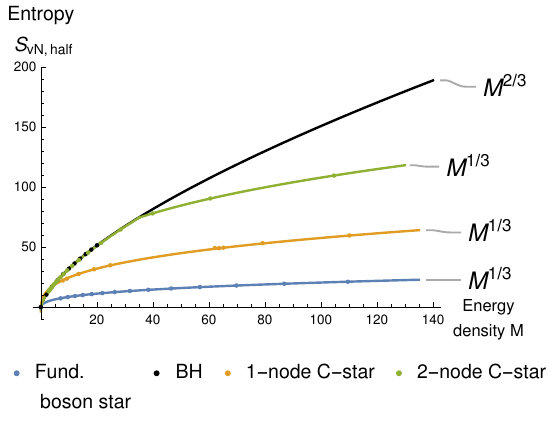}
    \caption{Asymptotically $AdS_4$: boundary EE of half the system for extremal RN black hole, C-star and fundamental boson star in the strong coupling regime when $e^{\rm eff}=3>e^{\rm eff}_{\rm crit,2}$.
    Dots are numerical data, solid lines represent a fit.
    The corresponding  phase diagram is shown in Figure \ref{fig:postC}. }
    \label{fig:ads4_postcross}
\end{figure}
  
  \textbf{"Boring" weak coupling phase:} $e^{\rm eff}<e^{\rm eff}_{\rm crit,1}$:  In this case we can start from a normal mode solution $e_j(x)$ and increase the amplitude. It turns out that all solutions have a bounded mass: at first the mass grows with the amplitude $|\phi(0)|$, but then reaches the maximum and decreases. For amplitudes above the maximum of the mass the solution is linearly unstable. Technically the Figure \ref{fig:preC} illustrates the intermediate coupling phase, but the qualitative behavior of the normal modes is the same. Dashed blue (representing fundamental mode $e_0(x)$) and orange (first excited mode $e_1(x)$) shows the behavior of the frequency and ADM mass. Analytical arguments suggest \cite{Dias:2016pma} that $e^{\rm eff}_{\rm crit,1} = \sqrt{3/2}$. Our numerical results are consistent with this prediction, although the shooting becomes increasingly hard near the critical point.

\begin{figure}
    \centering
\includegraphics[scale=0.8]{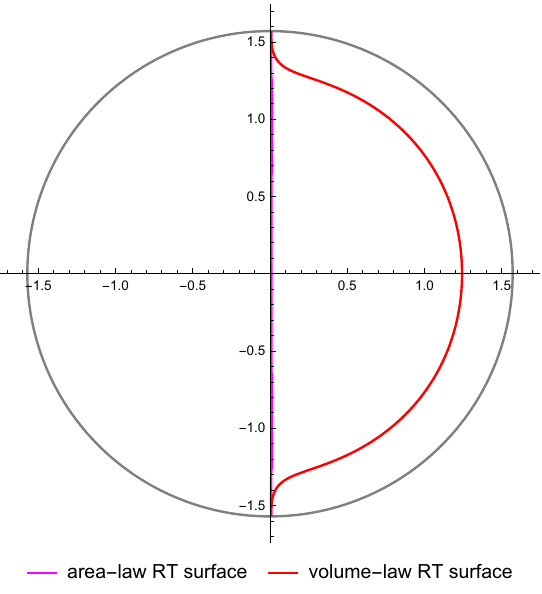}
    \caption{Time-slice $(x,\theta)$ of a C-star solution, $ds^2 = \frac{1}{\cos(x)^2} \l( A(x)^{-1} dx^2 + \sin(x)^2(d\theta^2 + \sin(\th)^2 d\varphi^2) \r)$ which is asymptotically global $AdS_4$ (each point contains a circle $\varphi$ which we suppressed). The gray circle indicates the conformal boundary at $x=\pi/2$.   }
    \label{fig:rt_compete}
\end{figure}

    \textbf{ Intermediate coupling:} $e^{\rm eff}_{\rm crit,1}<e^{\rm eff}<e^{\rm eff}_{\rm crit,2}$, illustrated by Figure \ref{fig:preC}.
    In this regime the solutions connected to the perturbative normal modes $e_j(x)$ behave qualitatively similar (dashed lines). 
    Interestingly, above certain amplitude additional solutions appear (solid lines).
    These solutions can have different number of zeros. The one with no zeros (solid blue) is usually called "the second branch of the fundamental mode" in the literature \cite{Gentle:2011kv,Dias:2011tj,Arias:2016aig}. Why it is called the second branch will become apparent from its behavior in the strong coupling phase. 
    Take the fundamental second branch with no zeros (solid blue).
 This branch has unbounded mass, but is linearly stable. Does it signal the presence of a scar? 
    Our answer for this question is: probably not. This state has a \textit{lower} mass compared to the extremal Reissner--Nordstr\"om (RN) black hole of the same mass \cite{Arias:2016aig}, it satisfies \cite{delaFuente:2020yua} the relation $M \sim Q^{3/2}$ expected for the ground state of a CFT. Moreover, it is horizonless, hence it is dual to a pure state of the boundary theory. Hence, we can expect that it is actually dual to (or at least very close to) a ground state of the CFT, as was proposed in \cite{delaFuente:2020yua}.
    Below we will also show that it has area law entanglement.
    Interestingly, we find solutions with unbounded mass which has more than one zero in the scalar profile: 1-node (solid orange) and  2-node (green) in the Figure \ref{fig:preC}, although they almost coincide. However, we will abstain from calling them "the second branch of excited modes". Instead, we call them "C-stars". Again, the reason for this will become clear from the strong coupling behavior. 
    They are significantly heavier than the fundamental (0-node) solution, so they are not close to the ground
    state at a fixed charge.
    \textit{We claim that these C-stars are approximate scar states.} Approximate means that they are not
    exact energy eigenstates, as discussed in the Introduction.
    To backup this statement, we evaluated the entanglement entropy of half the system for different values of $M$ - Figure \ref{fig:ads4_precross}. As we explained in Section \ref{sec:ee}, this probes the entanglement structure in the infinite-volume limit. We indeed find area-law entanglement $M^{(d-2)/d} = M^{1/3}$, in contrast to the volume law $M^{(d-1)/d} = M^{2/3}$ of the extremal RN black hole. Actually, depending on the mass, the relevant RT surface can have different configurations - Figure \ref{fig:rt_compete}.
    There is a simple RT surface (pink line) slicing through the equatorial plane, which yields area-law entanglement $M^{1/3}$. We found that it always dominates for large enough $M$, for both C-stars and the second branch of the fundamental mode. However,
    there is another RT surface which avoids the strong gravity region by curving around it (red line).
    For some C-stars, it dominates if $M$ is not too large, resulting in a entanglement shadow \cite{Czech:2012bh,Balasubramanian:2014sra,Freivogel:2014lja}: an area in the bulk which cannot be probed
    with extremal surfaces.

    The planar limit of a heavy fundamental boson star should coincide with zero temperature limit of a holographic superconductor \cite{Horowitz:2009ij}. One can verify independently that those have sub-volume law entanglement. 
    However, they seem to posses an interesting phenomena that at intermediate distances there is volume law scaling (c.f. Figures \ref{fig:ads4_precross},\ref{fig:ads4_postcross}). It would be interesting to understand this entanglement pattern from the boundary perspective.

    Our numerics suggests $e^{\rm eff}_{\rm crit,2} \approx 2.3$.

\begin{figure}
    \centering
    \includegraphics[scale=0.85]{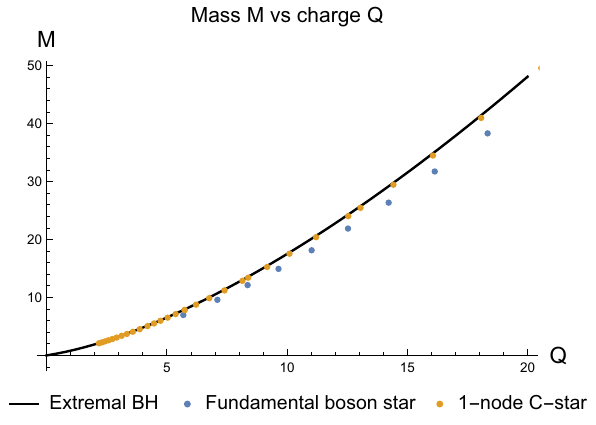}
    \caption{Mass $M$ vs charge $Q$ in the strong coupling regime $e_{\rm eff}>e_{\rm eff,2}$. 1-node C-star solution lies slightly below the extremal RN curve.}
    \label{fig:MvsQ}
\end{figure}

    \textbf{Strong coupling:} $e^{\rm eff}>e^{\rm eff}_{\rm crit,2}$. In this case the two branches of the fundamental mode merge: one can start from small-amplitude normal mode solution and monotonically increase mass to infinity - Figure \ref{fig:postC} (blue). This is the origin of the name "second branch" in the intermediate coupling phase. In contrast, this does not happen (at least for $e^{\rm eff} \le 20$, because we are limited by numerics) to the excited modes and C-stars - Figure \ref{fig:postC} (orange, green). This is why we do not call C-stars "the second branch of the excited modes".
    Both C-stars and the fundamental boson star are superextremal - Figure \ref{fig:MvsQ} and have area-law entanglement - Figure \ref{fig:ads4_postcross}. 
    So they continue being scar states at strong coupling.

One important question is the stability of the solutions we found.
This question is important because unstable solutions can be highly sensitive to the parameters of the theory. For example, in this paper we neglected a possible explicit self-interaction of the scalar field (apart from the one mediated by gauge field and gravity). In holographic theories we expect the bulk fields interactions to be small, but non-zero.
The first step to understand boson star stability is to consider normal modes of the linearized boson star perturbations. We follow the method outlined by \cite{Arias:2016aig}, which reduces the system of equations for spherically symmetric linearized perturbations of a boson star to a system of three linear equations, two for real and imaginary scalar field perturbations and one for gauge potential perturbations, and then uses Chebyshev pseudospectral collocation method to find normal modes of these equations. The star is stable when all the modes have real frequencies and becomes unstable when at least one of the frequencies becomes imaginary. The point of transition between the stable and the unstable parts of the branches of boson stars happens when the mass curve encounters an extremum $M'(\varepsilon) = 0$, there is a good argument to that \cite{Arias:2016aig} that follows a well-known argument for fluid stars \cite{Weinberg:1972kfs} and it stays valid for all the AdS boson stars we have encountered. Furthermore, linearly stable boson stars in AdS are known to be non-linearly stable as well \cite{Buchel:2013uba, Maliborski:2013jca, Arias:2016aig}. Figures \ref{fig:preC} and \ref{fig:postC} show the square of the frequency of the lowest normal mode, all the parts of the branches with unbounded mass (separated by a mass extremum) are stable.
Because they are linearly stable, we do not expect that self-interaction or higher-derivative curvature terms will affect the solutions.
Direct studies of fundamental boson stars in various theories  \cite{PhysRevLett.57.2485,PhysRevD.38.2376,Balakrishna:1997ej,PhysRevD.42.384,Schunck:1999zu,Sanchis-Gual:2021phr,Buchel:2013uba,Brihaye:2014bqa,Hartmann:2013tca,Henderson:2014dwa} support this intuition.

\section{State preparation}
\label{sec:prep}
We have identified a bulk field configuration which has scar properties. How do we prepare this state using CFT path integral? The calculations we presented above concerned purely classical system of Einstein gravity with a scalar field. It means that in the QFT language we are dealing with a coherent state of that scalar field. The question of preparing such coherent states was addressed in a series of papers \cite{Skenderis:2008dg,Skenderis:2008dh,Botta-Cantcheff:2015sav,Christodoulou:2016nej,Faulkner:2017tkh,Marolf:2017kvq}.

Since we want a single-mode configuration at $t=0$, which is spherically symmetric, we prepare CFT state on a sphere using disk (hemisphere) path integral, where we insert a single $e^{\tilde{\vep} \Oc}$ operator at the center of the disk - Figure \ref{fig:eucl_prep}.
$\Oc$ is CFT single-trace operator dual to the field $\phi$ in the bulk. For $\mathcal{N}=4$ super Yang--Mills such operator could be $\Tr F^2$ or $\Tr \tilde{F} F$ ($F$ being the gauge field strength) which is marginal and has no $R-$charge in order to leave the $S^5$ part of the bulk geometry intact.
Constant $\tilde{\vep}$ is proportional (up to an order-one number) to the bulk scalar field amplitude $\vep$. If the two-point function of $\Oc$ is normalized to $1$, then $\tilde{\vep}$ can be large, of order the square-root central charge, such that gravity backreaction becomes important. This state is fine-tuned: moving the operator away from the pole will create a spherically non-symmetric configuration which will collapse to a black hole at times of order
$1/\vep^2$.

In order to read off the bulk configuration we need
bulk-to-boundary propagator which for global $AdS$ is
\beq
K(x,\hat{e},t; \hat{e'},t') = 
\l( \frac{\cos(x)}{\cos(t-t')+\sin(x) (\hat{e},\hat{e'})} \r)^{\Delta},
\eeq
where unit vector $\hat{e}$ parameterize $S^{d-1}$ of $AdS_{d+1}$.
Also it is important to keep in mind that we are interested in expectation values in the form 
$\bra \Psi | \cdot | \Psi \ket$, so the path integral involves both south and north hemispheres.
Putting the operators at the poles basically \footnote{This involves an auxiliary conformal transformation. For example, for $AdS_3$, one introduces $z=e^{i t' + i l'}, \bar{z}=e^{i t'-i l'}$, where $l'$ is angle in the equatorial plane. 
Center of the disk is $z=\bar{z}=0$.
The propagator looks like
\beq
K = \l( \frac{2 \cos(x) \sqrt{z \bar{z}} }{e^{it}+e^{-it} z \bar{z} + \sin(x) [e^{il} \bar{z}+e^{-il} z]} \r)^{\Delta}.
\eeq
} sets $(\hat{e},\hat{e}')=0$. This yields $\phi \propto \tilde{\vep} \cos(x)^{\Delta}, \pr_t \phi=0$ profile at $t=0$, which is what we want for an oscillon. 
In case of a complex field, we get $\phi \propto \tilde{\vep} \cos(x)^{\Delta}, \pr_t \phi \propto i \Delta \tilde{\vep} \cos(x)^{\Delta}$, which is the relevant configuration for a boson star. 
Excitations with higher radial numbers can be obtained by acting with derivatives. For example, for the first excited mode we need to insert $\pr^2 \Oc(0)$, $\pr^2$ being the Laplacian on the sphere.
In a generic quantum field theory an insertion in the form $e^{\tilde{\vep} \Oc(0)}$ is not well-defined. Thanks to the stability of boson stars/oscillons
one can introduce a small smearing 
$e^{\tilde{\vep} \int d^d z \Oc}$. Such operator is well-defined and it would be interesting to investigate whether such operators lead to scars beyond holographic CFTs.

Of course, this is just a leading order in $\tilde{\vep}$ (that is, in $\vep$) answer. One can solve bulk equations of motions perturbatively and then prepare the configuration at $t=0$ exactly by placing the appropriate Euclidean sources. We refer to  \cite{Marolf:2017kvq} for a discussion.

\section{Volume-law entanglement and weak energy conditions}
\label{sec:weak_energy}
In the previous sections we demonstrated that oscillons have smaller entanglement
entropy compared to black holes and boson stars have
parametrically smaller entanglement entropy compared to a black hole of the same mass. Namely, for half-system the black hole answer is "volume-law" $M^{(d-1)/d}$, whereas for C-stars it is "area-law" $M^{(d-2)/d}$.

 In this Section we show that imposing the weak-energy condition in the bulk guarantees that the CFT entanglement entropy is smaller compared to the one of a thermal state of the same mass, which is a well--known statement from the statistical mechanics. Unfortunately,
 our arguments are not sensitive to the presence/absence of the horizon. It would be interesting to understand how the requirement
 of horizon absence further bounds the entanglement.

Consider a static space-time
(\ref{eq:geon_metric}) with some matter fields, with or without a horizon. Weak energy condition $T_{tt} \ge 0$ for the matter stress-energy tensor yields
\beq
\pr_x A \le \frac{d-2+2 \sin^2(x)}{\sin(x)\cos(x)} (1-A).
\eeq
Gr\"onwall theorem implies\footnote{Specifically we need to apply it for the interval $[0,\pi/2)$ in the vicinity of $x=\pi/2$ to connect constant $M$ with spacetime mass.} that $A$ can be bounded by the solution of the corresponding differential equation:  
\beq
\label{eq:Abound}
A \ge 1 - 2 M \frac{\cos(x)^d}{\sin(x)^{d-2}}.
\eeq
The right hand side is the value of $A$ for AdS black hole of mass $M$. In particular, it implies that if there is a horizon, it lies inside a would-be black hole of the same mass\footnote{We can transfer surfaces from one geometry to another because we gauge-fixed the metric to be in the form (\ref{eq:geon_metric}).}. Imagine now, that we fix a boundary subregion and the energy density $M$. Then the RT prescription in the corresponding black hole background with this mass will yield a certain co-dimension 2 surface. Now, if we compute the area of the same surface in the geometry of interest of the same energy density, the area will be lower because of the inequality (\ref{eq:Abound}). But the RT prescription instructs us to find the minimum over all possible surfaces, so the actual RT answer will be even lower.

\section{Conclusion}
In this paper we studied the properties of oscillons and boson stars in asymptotically $AdS$ spacetime. These are time-periodic, horizonless, solitonic solutions and we argued that they are linearly stable, have low-entanglement and are easily prepared with the Euclidean path integral. In the dual CFT they signal the presence of scars states. We demonstrated that excited boson
stars have area law entanglement. In contrast, in low dimensional spin-chains scar states often have logarithmic scaling of entanglement. It is an interesting question if it is possible to obtain something like this in holographic theories in higher dimensions. Also it would be interesting to go beyond entanglement and understand other properties of scar states, both holographic and not. For example, entanglement can be used as a probe for
confinement \cite{Klebanov:2007ws}.

Oscillons and boson stars only require the presence of a scalar field in a theory, hence they represent a generic phenomenon for holographic theories. As we mentioned in the Introduction, one can even build solitonic objects (geons) purely from the gravitational degrees of freedom \cite{Horowitz:2014hja}. It would be interesting to perform the analysis of this paper for geons.

There are two important take-aways, one for the holographic CFTs and one for the gravity.

The gravity predicts that for holographic CFTs the states corresponding to $e^{\tilde{\vep} \Oc}$ (prepared by Figure \ref{fig:eucl_prep}) are non-thermalizing finite energy-density states. It means that in any holographic CFT, 4-point correlation function of the form 
\beq
\label{eq:4pt}
\bra e^{\tilde{\vep} \Oc} L L e^{\tilde{\vep} \Oc} \ket,
\eeq
where $L$ are light fields evolved in Lorentzian signature, and $e^{\tilde{\vep} \Oc}$ are inserted at the poles of the Euclidean sphere,
will not look thermal in any number of dimensions. It is important to emphasize that  $\tilde{\vep}$ can be large, of the order of square-root central charge (if $\bra \Oc \Oc \ket$ is normalized to 1), so this operator causes huge backreaction.
In contrast, states prepared with the insertion finitely away from the pole will look thermal.
The question of thermality of CFT correlators has been extensively studied before. 
For the case of heavy conformal primaries $H$ in 2d CFT, it was argued in \cite{Fitzpatrick:2015zha,Fitzpatrick:2016ive} that 4-point function of the form
\beq
\bra  H L L H \ket,
\eeq
does look thermal.
Also \cite{Anous:2016kss, Anous:2017tza}  studied more complicated CFT states which effectively prepare collapsing dust shells in the dual gravity. In both cases, the large central charge limit and the vacuum dominance in the 4-point function was enough to link the CFT results to the gravity black hole computation.
It would be interesting to perform a similar CFT calculation for the (\ref{eq:4pt}) correlator and map the results to the oscillon (boson star) background.

On the other hand, all current examples of scar states in the condensed matter literature involve the presence of hidden symmetries \cite{Pakrouski:2020hym,Lin_2019,Mark_2020,Ren:2020dcs,Pakrouski:2021jon,Ren:2021khc,Moudgalya:2022nll,Sun:2022oew}.
A possible hidden symmetry in behind oscillons was discussed in \cite{Buchel:2014xwa,Evnin:2015gma, Cardoso:2017qmj, Craps:2014vaa}. 
Without a backreaction, a scalar field
of mass $m^2 = \Delta(\Delta-d)$ in $AdS_{d+1}$ posses a set of normal modes:
\beq
\omega_{jl} = \Delta + l + 2j, \  j,l=0,1,\dots, 
\eeq
where $l$ is the (integer) angular momentum and $j$ is the (integer) radial quantum number. Notice that they enter only in $l+2j$ combination. This highly resonant spectrum is a direct consequence of the conformal $SO(2,d)$ symmetry of $AdS$. Such resonant spectrum is the reason why $AdS$ is unstable, because non-linearities coming from gravity may produce secular corrections growing linearly in time. Surprisingly, it was argued in \cite{Evnin:2015wyi} that the same $SO(2,d)$ symmetry forces
the secular terms to vanish. 
This symmetry constrains a lot the leading non-linear correction to the oscillon solution, but it would be interesting to understand what role this possibly weakly-broken symmetry plays for the full non-linear solution.

As we mentioned many times, oscillons and boson stars are only approximate energy eigenstates which only signal the presence of scar energy eigenstates in the spectrum.
Is it is possible to construct a geometry dual to the actual scar eigenstate? One possible set of candidates are Lin--Lunin--Maldacena (LLM) \cite{Lin:2004nb} geometries.
They are dual to half-BPS boundary operators \cite{Corley:2001zk,Berenstein:2004kk}. 
These geometries are complicated, but it would be nice to understand what entanglement law they have, some preliminary steps in this direction were made in \cite{Balasubramanian:2017hgy}.
However, LLM solutions are very special: they preserve supersymmetry and explicitly use the compact $S^5$ part of the geometry. Unlike oscillons and boson stars, we do not expect to find something like this in more generic holographic theories.

\section*{Acknowledgment}
We would like to thank F.~Popov and K.~Pakrouski for
for collaboration at the early states.
We are grateful to
D.~Berenstein, A.~Buchel, E.~Colafranceschi, S.~Colin-Ellerin, A.~Dymarsky, A.~Gorsky, A.~Holguin, L.~Lehner, D.~Marolf, F.~Pretorius, M.~Srednicki, A.~Zhiboedov
and specially to S.~Hellerman, G.~Horowitz, I.~Klebanov for comments and discussions.
 This material is based upon work supported by the Air Force
Office of Scientific Research under award number FA9550-19-1-0360. It was also supported
in part by funds from the University of California. 
AM would like to thank 
C.~King for moral support.
%\printbibliography
\bibliography{refs}

\begin{thebibliography}{121}
\bibitem [{Notea()}]{Notea}%
  \BibitemOpen
  \bibinfo {note} {We are indebted to F.~Popov for asking the first question
  and explaining to us the difference between the various scar
  phenomena.}\BibitemShut {Stop}%
\bibitem [{Noteb()}]{Noteb}%
  \BibitemOpen
  \bibinfo {note} {Einstein--Gauss--Bonnet gravity has a mass gap for black
  holes, making $AdS$ stable again. The claim is that there are initial
  conditions, above this gap, which do not lead to black hole
  formation}\BibitemShut {NoStop}%
\bibitem [{Noted()}]{Noted}%
  \BibitemOpen
  \bibinfo {note} {This involves an auxiliary conformal transformation. For
  example, for $AdS_3$, one introduces $z=e^{i t' + i l'}, \protect \overline
  {z}=e^{i t'-i l'}$, where $l'$ is angle in the equatorial plane. Center of
  the disk is $z=\protect \overline {z}=0$. The propagator looks like \begin
  {equation}K = \left ( \protect \frac {2 \cos (x) \protect \sqrt {z \protect
  \overline {z}} }{e^{it}+e^{-it} z \protect \overline {z} + \sin (x) [e^{il}
  \protect \overline {z}+e^{-il} z]} \right )^{\Delta }. \end
  {equation}}\BibitemShut {NoStop}%
\bibitem [{Notee()}]{Notee}%
  \BibitemOpen
  \bibinfo {note} {Specifically we need to apply it for the interval $[0,\pi
  /2)$ in the vicinity of $x=\pi /2$ to connect constant $M$ with spacetime
  mass.}\BibitemShut {Stop}%
\bibitem [{Notef()}]{Notef}%
  \BibitemOpen
  \bibinfo {note} {We can transfer surfaces from one geometry to another
  because we gauge-fixed the metric to be in the form (\ref
  {eq:geon_metric}).}\BibitemShut {Stop}%
\end{thebibliography}
\end{document}